\documentclass[5p]{elsarticle}
\usepackage{textcomp}
\usepackage{fleqn}
\usepackage{graphicx}
\usepackage{hyperref}
\usepackage{pifont}
\usepackage{amsmath,amssymb}
\usepackage{natbib}
\usepackage{miller}

\usepackage[utf8]{inputenc}

\usepackage{xspace}

\newcommand{\etal}{\textit{et al}.\@\xspace}
\newcommand{\abinitio}{\textit{ab initio}\@\xspace}

\newcommand{\ie}{\textit{i.e.}\@\xspace}

\begin{document}

\title{Influence of simple metals on the stability of $\hkl<a>$ basal screw dislocations in hexagonal titanium alloys}

\author[1]{Piotr Kwasniak\corref{cor1}}
\ead{piotr.kwasniak@pw.edu.pl}
\author[2]{Emmanuel Clouet}
\address[1]{Faculty of Materials Science and Engineering, Warsaw University of Technology, Woloska 141, 02-507 Warsaw, Poland}
\address[2]{DEN-Service de Recherches de Métallurgie Physique, CEA, Université Paris-Saclay, F-91191 Gif-sur-Yvette, France}
\cortext[cor1]{Corresponding author}

\begin{abstract}

Basal slip acts as a secondary deformation mode in hexagonal close-packed titanium
and becomes one of the primary mechanisms in titanium alloyed with simple metals.
As these solute elements also lead to a pronounced reduction of the energy of the basal stacking fault, 
one can hypothesize that they promote basal dissociation of dislocations which can then easily glide in the basal planes.
Here, we verify the validity of this hypothesis using \abinitio calculations
to model the interaction of a screw dislocation with indium (In) and tin (Sn).
These calculations confirm that these simple metals are attracted by the stacking fault existing in the dislocation core 
when it is dissociated in a basal plane,
but this interaction is not strong enough to stabilize a planar configuration,
even for a high solute concentration in the core. 
Energy barrier calculations reveal that basal slip, in the presence of In and Sn, proceeds without any planar dissociation, 
with the dislocation being spread in pyramidal and prismatic planes during basal slip like in pure Ti.
The corresponding energy barrier is higher in presence of solute atoms, 
showing that In and Sn do not ease basal slip but increase the corresponding lattice friction. 
This strengthening of basal slip by solute atoms is discussed in view of available experimental data.

\end{abstract}


\maketitle

\section{Introduction}

Plastic deformation of hexagonal close-packed (hcp) titanium is realized through dislocation glide
in various slip systems and through twinning.
The activity of these different deformation modes strong\-ly depends on the temperature and material purity \cite{Churchman1954,Conrad1981,Naka1982,Biget1989,Williams2007}. 
At low temperatures, usually below $\sim550$\,K, 
prismatic slip of $\hkl<a>=1/3\,\hkl<1-210>$ dislocations is the dominant deformation mode.
As the temperature increases, basal and pyramidal slip of the same \hkl<a> dislocations becomes more active. 
Glide of \hkl<c+a> dislocations or twinning can also be activated to accommodate strain along the \hkl<c> axis 
of the hcp lattice, but the corresponding critical resolved shear stresses (CRSS) 
remain few times higher than the CRSS of \hkl<a> slip modes, even at elevated temperatures \cite{Williams2007,Gong2009,Barkia2015}.

Like other hcp metals, titanium therefore experiences plastic anisotropy which can be detrimental to its ductility 
and formability. 
Activity of different slip modes can be, however, modified by alloying elements, opening a way to reduce plastic anisotropy.
One good example is titanium alloyed with simple metals like aluminum \cite{Truax1974,Boyer1994,Williams2007},
as the strengthening caused by Al solute varies among the active deformation modes.
Measurements in Ti-Al single crystals show that the ratio of the CRSS 
of basal and prismatic slips decreases with Al addition \cite{Sakai1974,Williams2002}.
As a result, prismatic slip is favoured at room temperature in pure Ti, 
but the stresses necessary to activate basal and prismatic slip become essentially equal 
in titanium with 6.6\,wt.\% Al over a wide range of temperatures \cite{Williams2002}.
This evolution of the mechanical properties goes with a change of the dislocation microstructure
as evidenced by transmission electron microscopy (TEM). 
While the microstructure is homogeneous at low Al concentration, it becomes highly localised
when the Al content increases, with the development of planar slip bands 
whose thickness depends on the temperature and the Al concentration \cite{Truax1974,Williams2002}.
The above changes in deformation modes and mechanical properties are not limited to Ti-Al alloys 
but also exist in titanium alloyed with other simple metals like Sn or Ga
\cite{Collings1975,Zaefferer2003,Li2013,Qiu2014,Ye2017}. 

In these $\alpha$-Ti alloys, the activity of \hkl<a> slip systems is controlled 
by the motion of the \hkl<a> screw dislocations which experience a larger lattice friction 
than orientations with an edge character. 
This is evidenced by TEM observations which generally show long straight dislocations 
aligned along their screw orientation while gliding in the prismatic planes 
\cite{Naka1988,Farenc1993,Yu2015,Barkia2017}, the principal slip system,
or in the basal and pyramidal planes \cite{Caillard2018},
the secondary slip systems.
The same long screw dislocations are observed in presence of simple metals like Al
\cite{Williams2002,Sakai1974,Sakai1974b,Neeraj2000,Ambard2001,Castany2007}.
At low temperatures, the lattice friction opposing prismatic glide 
has been rationalized by the existence of a sessile ground state of the screw dislocation
which is dissociated in a pyramidal plane and
needs to transit to a metastable state of higher energy to glide in the prismatic planes \cite{Farenc1995,Clouet2015}.
Above room temperatures, where the friction associated with this locking-unlocking glide mechanism 
becomes negligible, the lattice friction originates from the interaction of interstitial solute elements, 
in particular oxygen which is inevitably present in titanium alloys, with the core of the screw dislocations
\cite{Naka1988,Yu2015,Chaari2019}.
As for secondary slip systems, 
\abinitio calculations have shown that \hkl<a> screw dislocations gliding in pyramidal
or basal planes need to overcome a high energy barrier \cite{Clouet2015,Kwasniak2019},
thus leading to a mobility controlled by the nucleation and propagation of kink pairs in both cases 
\cite{Clouet2015,Caillard2018}.

Understanding the impact of simple metals on the activity of the different slip systems in titanium 
therefore goes through the study of the interaction of these solute atoms with 
\hkl<a> screw dislocations.  
Previous \abinitio calculations have shown that simple metals lead in titanium 
to energy variations of the stacking faults controlling dissociation of 
\hkl<a> dislocations in hcp metals \cite{Kwasniak2016}. 
In particular, a strong decrease of the fault energy is obtained for the basal stacking fault
while the energy of the prismatic and pyramidal faults either increases or only slightly varies. 
Such calculations thus indicate that an addition of simple metals 
eases the shearing of basal planes compared to prismatic and pyramidal planes.
It is then appealing to correlate this variation of the stacking fault energies with the decrease of the ratio
between the CRSS of basal and prismatic slips which has been experimentally observed 
with Al addition. 
But this assumption is nevertheless questionable as \abinitio calculations in pure Ti
have shown that a basal dissociation of the \hkl<a> screw dislocation is indeed unstable
and that the screw dislocation can glide in a basal plane while remaining dissociated 
in a prismatic or a pyramidal plane, thus without developing any basal stacking fault
\cite{Kwasniak2019}.
One can thus wonder if a basal dissociation of the screw dislocation 
can be stabilized by an addition of simple metals, 
thus offering an explanation to the decrease of the CRSS ratio between 
basal and prismatic slip.

In this article, we present the results of \abinitio calculations 
to clarify the influence of simple metals on the stability 
of an eventual basal dissociation of the \hkl<a> screw dislocations in hcp Ti alloys,
and the impact of these solute elements on basal glide of this dislocation.
The computations are performed for two Ti-based binary systems using Sn and In as alloying elements.
Sn is chosen as it leads to the largest reduction of the basal stacking fault energy
among all simple metals \cite{Kwasniak2016},
whereas In is chosen as its interaction with the \hkl<a> screw dislocation
promotes a compact core of this dislocation \cite{Kwasniak2017}, 
which may then equally glide in the different planes containing the \hkl<a> direction.

\section{Methodology}

\begin{figure}[ht!] 
\centering
\includegraphics[width=\columnwidth]{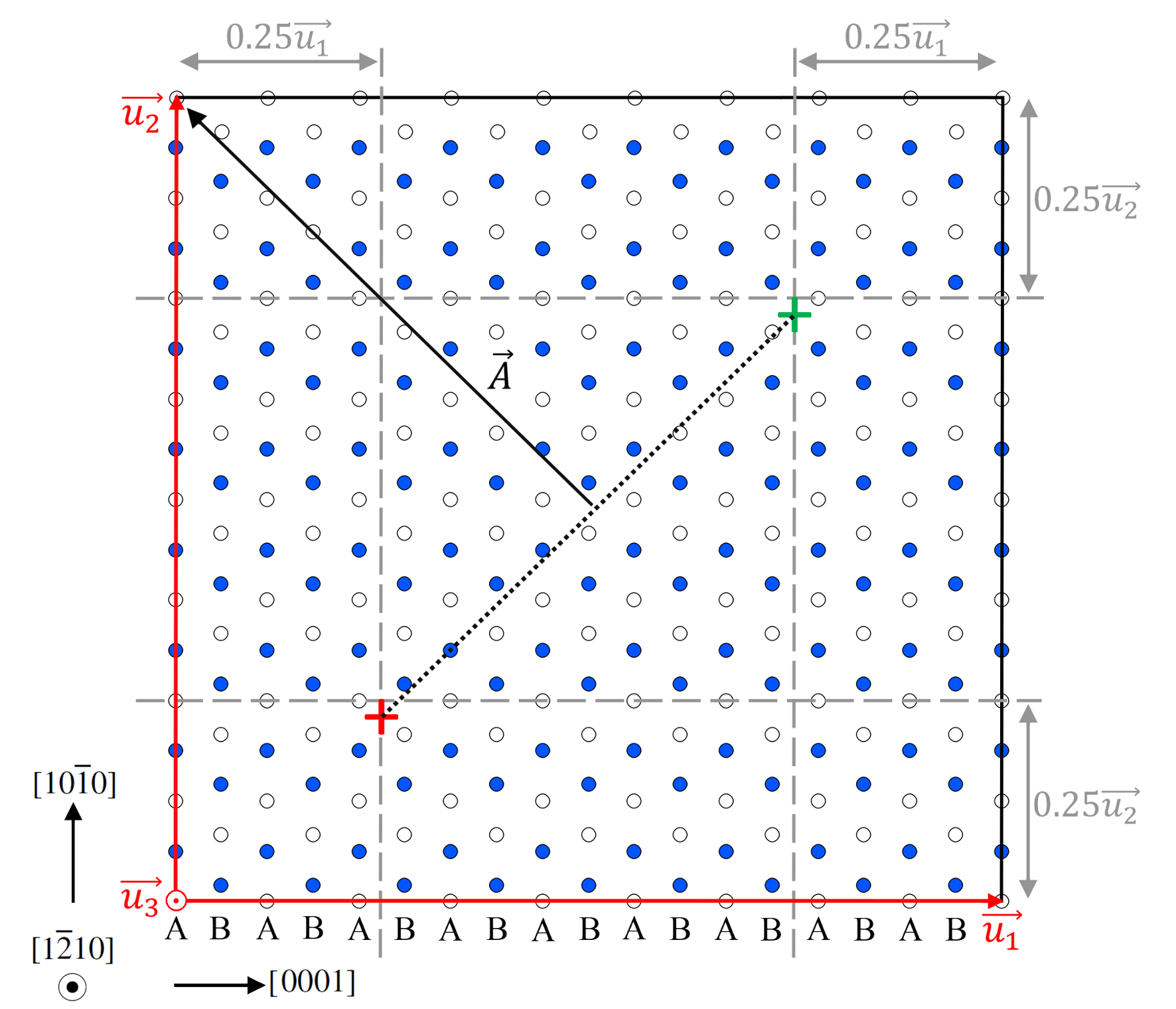}
\caption{Computational supercell with a quadrupolar arrangement of screw dislocations. Differently
coloured “+” symbols indicate the centre of dislocations with opposite Burgers vector, $\vec{A}$ is the dipole cut vector.
The A and B symbols represent the periodic stacking of the basal planes. 
Blue and white dots denote positions of atoms lying on subsequent \hkl(1-210) atomic planes.}
\label{Fig1}
\end{figure}

The computational supercell used in this study contains 576 atoms and exhibits \textit{m} = 9, \textit{n} = 8 and \textit{l} = 2 unit cell periodicity along 
$\vec{e}_x$ = [0001], $\vec{e}_y$ = [10$\bar{1}$0] and $\vec{e}_z$ = 1/3\,[1$\bar{2}$10] directions, respectively. As a result, the periodicity vectors of the simulation box are $\vec{u}_1 = mc\vec{e}_x$, $\vec{u}_2 = na\vec{e}_y$ and $\vec{u}_3 = la\vec{e}_z$,
where $a$ and $c$ are the Ti lattice parameters. 
Full periodic boundary conditions were used \cite{Clouet2012,Rodney2016} with a quadrupolar arrangement of screw dislocations, employing the same setup as in our previous study \cite{Kwasniak2019}, except that the length of the supercell has been doubled along the dislocation line ($l = 2$) to reduce the interaction of the solute atoms with their periodic images.
Structure of the simulation box together with position of dislocation cores are shown in Fig. \ref{Fig1}.
The same number of solute atoms are introduced, in equivalent positions, in the two dislocation cores composing the dipole.

\mbox{\textit{Ab initio}} calculations were performed with 
VASP code \cite{Kresse1993,Kresse1996}, with projector augmented wave (PAW) method 
for core-valence electron interaction \cite{Blchl1994} and Perdew-Burke-Ernzerhof (PBE) \cite{Perdew2016} 
generalized gradient functional.
The Brillouin zone was sampled in accordance with the \linebreak Monkhorst-Pack scheme \cite{Monkhorst1976} using $1\times1\times5$ gamma centered k-points grid and a 0.3\,eV Methfessel-Paxton electronic occupancy smearing. 
The Ti ($3d^2 4s^2$), Sn ($5s^2 5p^2$) and In ($5s^2 5p^1$) pseudopotentials were employed, with a 500\,eV cutoff energy for plane waves. 
We did not incorporate Ti $3p$ semi-core electrons in the valence state. 
As shown in the \ref{sec:pseudo}, these semi-core electrons have only a marginal contribution to the relative stability of the different core configurationas of screw dislocation in pure Ti.

The atomic structures of the screw dislocations were determined by performing 
ionic relaxations with a fixed-shape simulation box 
using a criterion of 3\,meV/\AA{} for Hellmann-Feynman forces convergence. 
For a dislocation interacting with $n_{sol}$ solute atoms (2 dislocations and $2 n_{sol}$ solute atoms in the simulation cell) the interaction energy between the dislocation and the solute atoms is defined as
\begin{equation} 
\Delta{}E = \frac{1}{2} \left(  E_{tot} - E_{\pi l} - n_{sol} E_{sol} + n_{sol} E_{Ti} \right),
\label{eq:Einter}
\end{equation}
where $n_{sol}$ is the number of solute atoms decorating each dislocation
and $E_{tot}$, $E_{\pi l}$, $E_{sol}$ and $E_{Ti}$ are the total energy of the same supercell 
which contains respectively a dislocation dipole with $n_{sol}$ solute atoms on each dislocation,
the dipole in pure Ti with the dislocations in their ground state (low energy pyramidal configuration \cite{Clouet2015}), 
two solute atoms in an otherwise perfect hcp Ti crystal at a separation distance large enough to prevent any interaction, and no defect nor solute.
As the basal configuration of the screw dislocation is unstable in pure Ti \cite{Kwasniak2019},
it is not possible to take this configuration as the reference state for the dislocation. 
We therefore took the dislocation ground state as a reference.
With such a definition, $\Delta E$ corresponds to the interaction energy between the dislocation and the solute atoms:
it describes the energy variation between a state where the dislocation is interacting with the alloying elements
and another state where the dislocation and the solute atoms are isolated,
with negative values of $\Delta E$ for attractive interaction.
When the solute atoms manage to stabilize the basal dissociation of the dislocation, 
$\Delta E$ contains a positive contribution corresponding to the energy difference between the two dislocation configurations.

\section{Stability of basal dissociation}

\subsection{Simulation setup}

We first examine if a screw dislocation can be stabilized by solute atoms in a configuration where it is dissociated in a basal plane.
Such a basal configuration has been found unstable in pure Ti, 
whatever the dissociation distance \cite{Kwasniak2019}.
The computations were performed for two positions of the dislocation center which are possible along the basal plane as shown in 
Figs. \ref{Fig2} and  \ref{Fig3}.
First position, marked as $B_1$, lies between two loosely spaced \hkl{10-10} prismatic planes and corresponds to the center of the ground state pyramidal ($\pi_l$) and of the metastable prismatic ($P$) dislocation cores, whereas the second position, $B_2$, is the center of the high energy pyramidal ($\pi_h$) core  located between two closely spaced \hkl{10-10} planes \cite{Clouet2015}.
The initial structures of the dislocations  were constructed according to anisotropic linear elasticity \cite{Babel}, for a dislocation dissociated in the basal plane in two 1/3\,\hkl[1-100] and 1/3\,\hkl[01-10] Shockley partials and homogeneous strain was added to the simulation box to compensate the plastic strain introduced by the dislocation dipole \cite{Rodney2016}.  
The dissociation distance of this initial configuration was taken equal to $\sim 15.3$\,{\AA}, a distance large enough to ensure that any possible stable basal configuration 
will be met during the reduction of the stacking fault ribbon which occurs while the configuration is relaxed.
Some additional calculations have been also performed for other initial configurations 
where we started from a smaller dissociation width for the initial state: 
the same configuration was obtained after relaxation.

According to recent \abinitio results of the \hkl<a> dislocations in Ti alloys, a single solute atom is not able to fully reconstruct the prismatic or pyramidal dislocation cores into the basal configuration \cite{Kwasniak2017}. 
Here, we go one step further, not only by starting from a configuration already dissociated in the basal plane,
but also by increasing the solute concentration in the stacking fault ribbon using a pair of solute atoms.
Additionally, dimension of the simulation box along the dislocation line is equal to two Burgers vector length to minimize the effect of solute self-interaction in this direction. 
Solute atoms are distributed symmetrically relative to the dislocation center, 
with one atom close to each partial dislocation.
The In and Sn atoms always occupy the next possible positions above and below the dissociation plane in the stacking fault region, since such sites leads to the strongest reduction of the  stacking fault energy \cite{Kwasniak2016}.
5 different separation distances $\lambda$ between the solute atoms in the \hkl[10-10] direction were studied for the two dislocation positions,
corresponding to the configurations $B_1(1-5)$ and $B_2(1-5)$ in Fig. \ref{Fig2} and Fig. \ref{Fig3}.
Solute atoms in the above configurations are located in the same  or first subsequent \hkl(1-210) atomic planes for $B_1(1-5)$ and $B_2(1-5)$, respectively.
Influence of solutes position along the dislocation line is discussed in \ref{sec:fluctuation} where it is shown to be limited.

\subsection{Relaxed configurations}

\subsubsection*{Dislocation center $B_1$}

\begin{figure*}[ht!]
\centering
\includegraphics[width=1\textwidth]{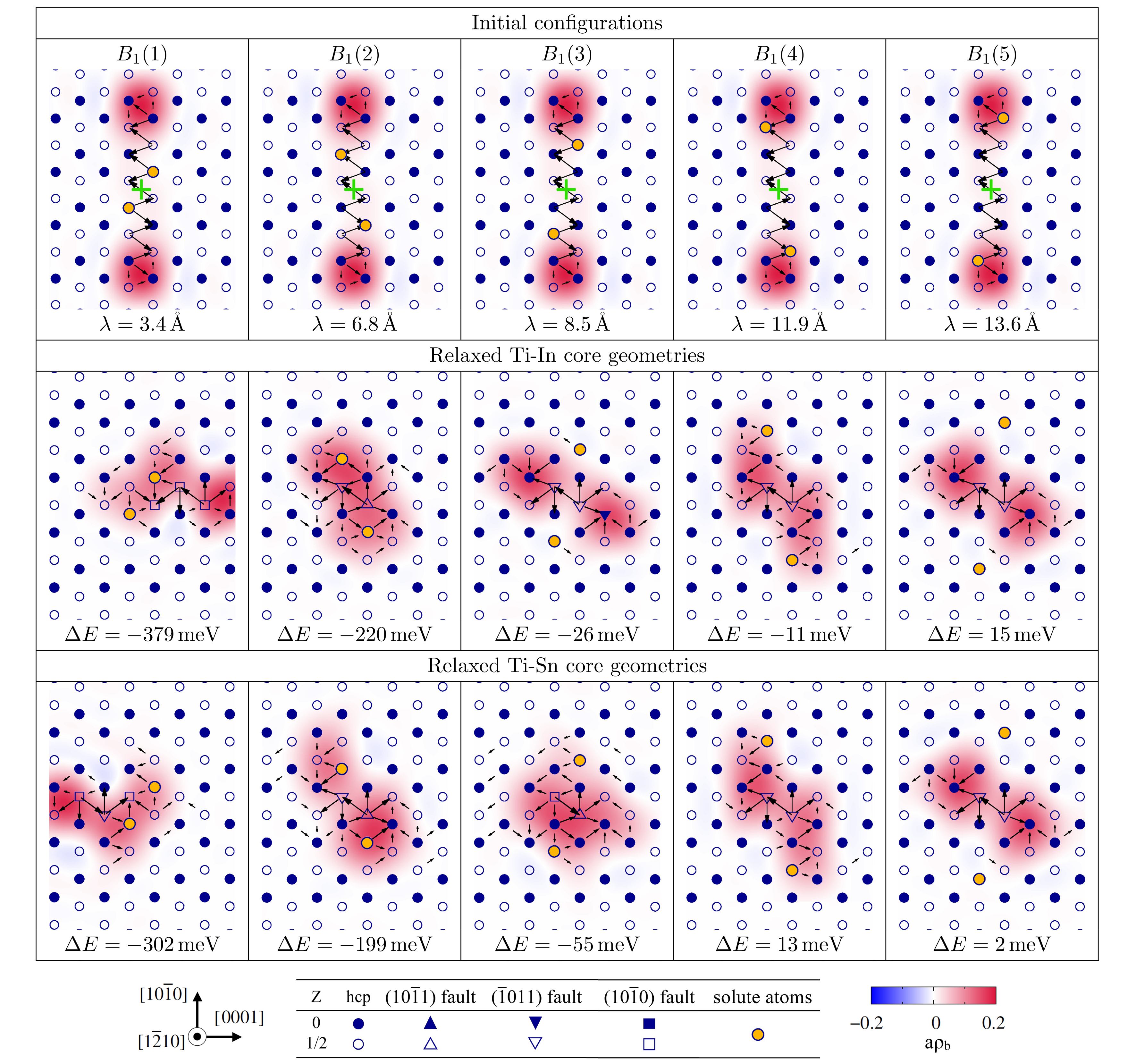}
\caption{Initial and relaxed core configurations $B_1(1-5)$ starting from a screw dislocation dissociated in the basal plane
and located at the position $B_1$ (green cross) for different separation distances $\lambda$
between solute atoms (orange circles) in Ti-In and Ti-Sn systems.
The arrows between two atomic columns are proportional to the differential displacement created by the dislocation in the [1$\bar{2}$10] direction. 
Displacements smaller than 0.1\,\textit{b} are not shown. 
The contour map shows the dislocation density according to the Nye tensor. 
Ti atoms belonging to particular types of stacking faults or different (1$\bar{2}$10) atomic planes are plotted with open or full-colored symbols as presented in the bottom of the Figure. }
\label{Fig2}
\end{figure*}

Initial and relaxed dislocation cores are shown in Fig.\,\ref{Fig2} for the dislocation position $B_1$.
None of these structures maintains its fully planar basal dissociation
and all relax to configurations which are spread in the pyramidal and prismatic planes. 
As solute atoms act as pinning points for the partial dislocations, some spreading of the core still exists in the basal plane,
with the amplitude of this basal spreading directly linked to the separation distance $\lambda$ between both solute atoms.
This pinning of the basal spreading is not efficient when the separation distance is too small (core $B_1(1)$) 
or too large (core $B_1(5)$). 
A configuration close to the pyramidal $\pi_l$ ground state or the metastable prismatic state in pure Ti is then obtained.
This is also the case with In for the intermediate distance $\lambda=8.5$\,{\AA} (core $B_1(3)$) where the dislocation core
relaxes to its ground state, whereas Sn manages to maintain basal spreading for the same separation distance.
For all other intermediate separation distances $\lambda$, relaxed structures are spanned between both solute atoms 
and spread on prismatic, pyramidal and basal planes, both with In and Sn.

\subsubsection*{Dislocation center $B_2$}

\begin{figure*}[ht!]
\centering
\includegraphics[width=1\textwidth]{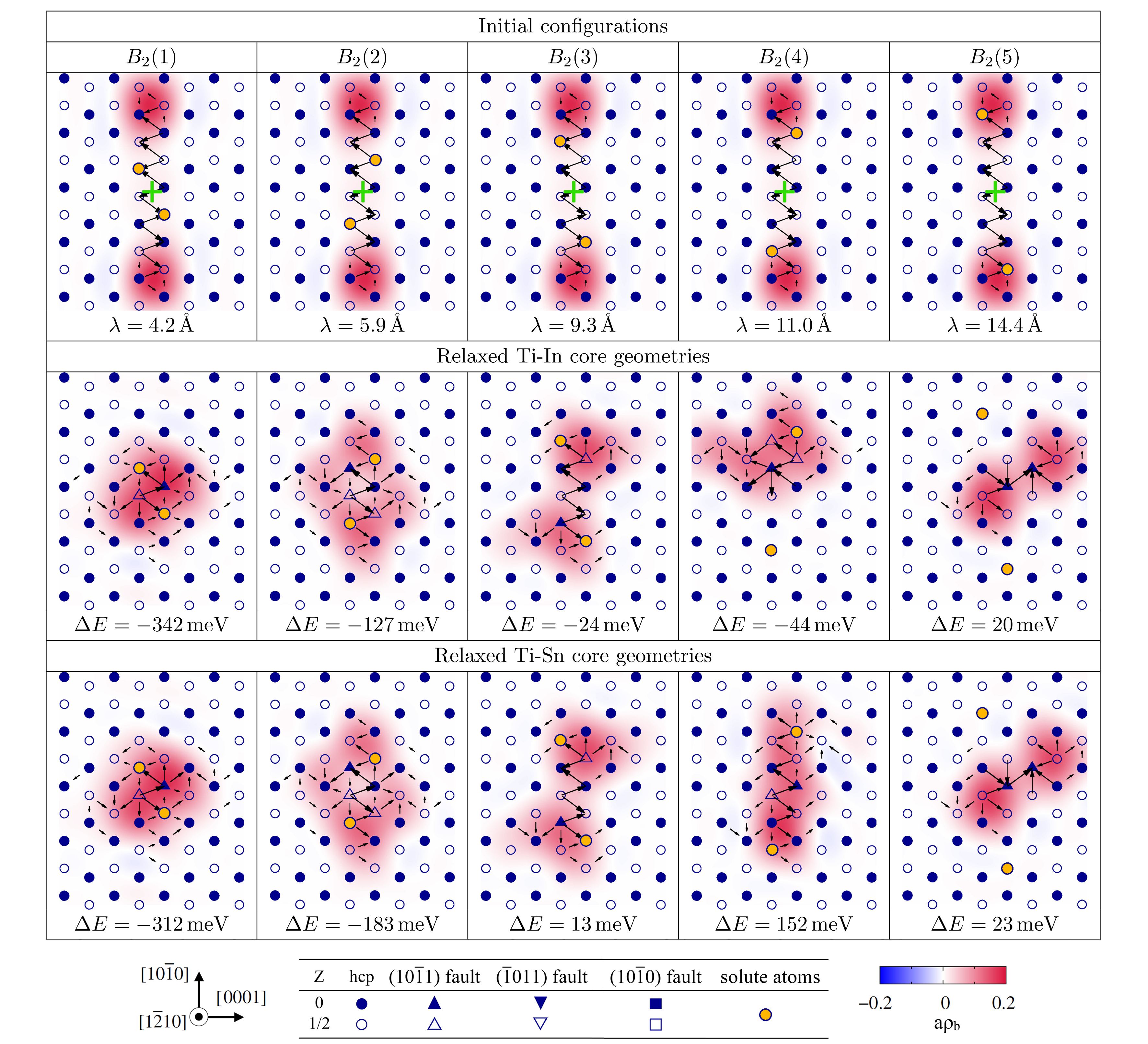}
\caption{Initial and relaxed core configurations $B_2(1-5)$ starting from a screw dislocation dissociated in the basal plane
and located at the position $B_2$ for different separation distances $\lambda$ between solute atoms in Ti-In and Ti-Sn systems.}
\label{Fig3}
\end{figure*}

We now examine relaxed structures obtained for the position $B_2$ of the dislocation (Fig. \ref{Fig3}).
Like for the $B_1$ position, solute atoms act as pinning points of the partial dislocations and enforce a basal spreading 
when their separation distance $\lambda$ is not too large.  
For $\lambda \geq 11$\,{\AA} with In and  $\lambda \geq 14$\,{\AA} with Sn, the pinning is not efficient anymore 
and the dislocation relaxes to a configuration dissociated in a pyramidal or prismatic planes. 
For closest pairs of solute atoms, the core is either compact (core $B_2(1)$ in Fig. \ref{Fig3})
or either non-planar with extended spreading on prismatic, pyramidal and basal planes.
Only with Sn for a separation distance $\lambda=11$\,{\AA} (core $B_2(4)$), an almost basal dissociation 
can be obtained, although some secondary spreading in the prismatic planes is still visible.

These calculations therefore show that In and Sn solute elements 
do not manage to stabilize in titanium a configuration of the screw dislocation 
with a planar dissociation in the basal plane.
Although some basal spreading can be seen for some positions of the solute atoms, 
additional prismatic or pyramidal spreadings still exist, leading to non-planar cores.

\subsection{Energy variation}

\begin{figure}[ht!]
\centering
\includegraphics[width=\columnwidth]{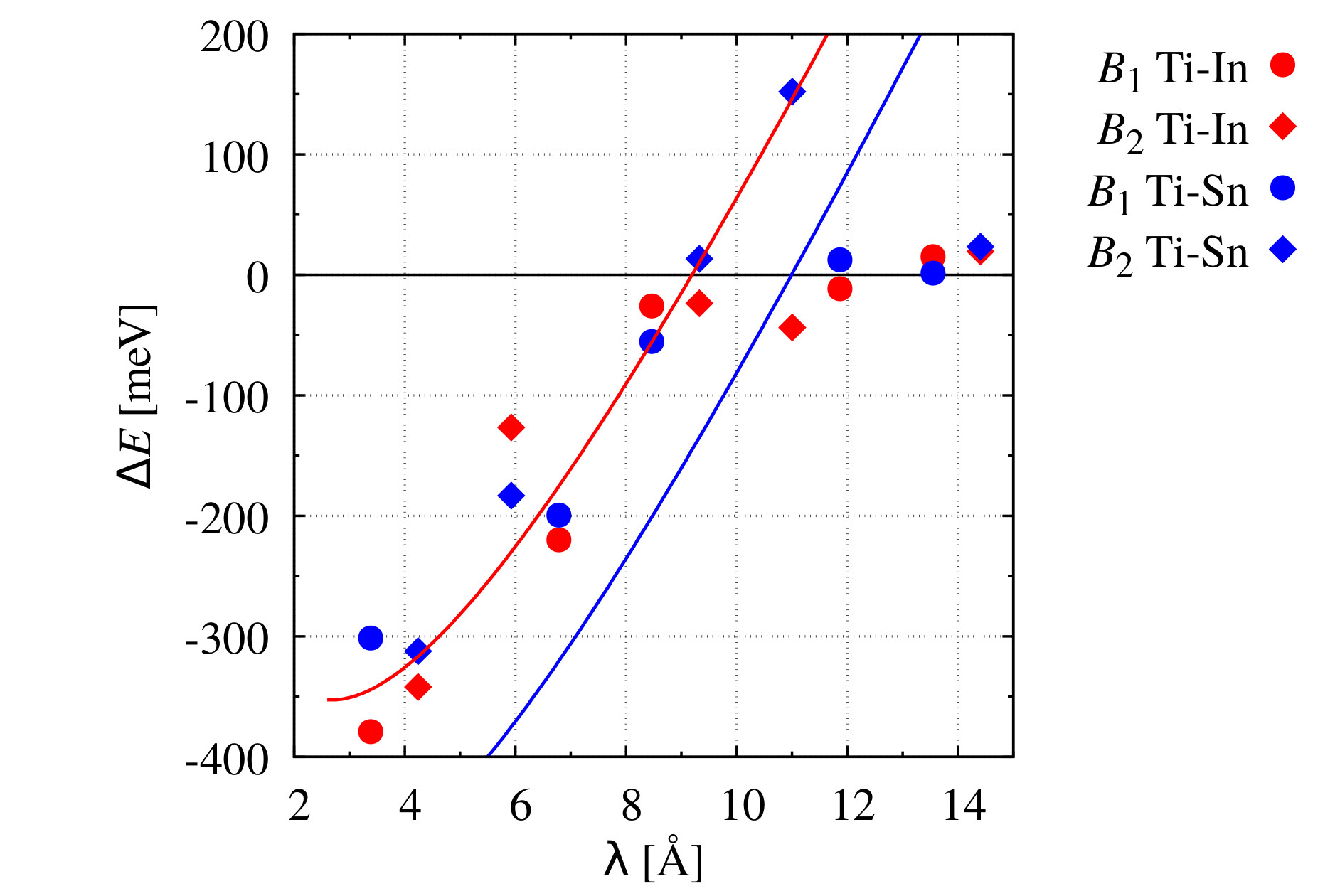}
\caption{
Interaction energy $\Delta E$ between the screw dislocation and the pair of solute atoms 
as a function of the separation distance $\lambda$ along the \hkl[10-10] direction between the two solute atoms.
Symbols are the results of \abinitio calculations (Eq. \ref{eq:Einter}) 
for the different positions of the solutes shown in Figs. \ref{Fig2} and \ref{Fig3}
and the solid lines are the predictions of the continuous model (Eq. \ref{eq:Einter_elas}).
}
\label{Fig4}
\end{figure}

The interaction energy $\Delta E$ between the screw dislocation and the pair of solute atoms is shown in Fig. \ref{Fig4}.
According to our definition (Eq. \ref{eq:Einter}), negative $\Delta E$ values correspond to attractive interaction.
The most stable configuration is obtained when the two solute atoms are at a minimal separation distance
$\lambda \sim 3$\,-\,$4$\,\AA.
This negative interaction energy then increases with the separation distance $\lambda$
until it becomes null for $\lambda$ larger than $\sim10$\,{\AA} where the solute atoms and the screw dislocation 
do not interact anymore.  
This is consistent with the relaxed structures shown in Figs. \ref{Fig2} and \ref{Fig3} 
where the screw dislocation adopts its ground state structure and is not perturbed
by the solute atoms for such large separation distances. 
The only clearly repulsive interaction is obtained for the configuration $B_2(4)$ in the Ti-Sn alloy:
this corresponds to the largest separation distance ($\lambda\sim11$\,\AA) for which the solute atoms manage
to maintain a basal spreading of the screw dislocation (Fig. \ref{Fig3}). 
For the same initial configuration in Ti-In, this basal spreading could not be stabilized 
and the dislocation relaxes to a core mainly spread in a prismatic plane, 
leading to a weak interaction.

For attractive configurations ($\lambda \lesssim 10$\,\AA), 
the main effect of the two solute atoms is to pin the Schockley partials 
and maintain the basal stacking fault ribbon between them.
Our previous study in pure Ti \cite{Kwasniak2019} has shown that the energy variations
with the dissociation distance of such a basal core is reasonably well described by linear elasticity. 
One thus expects that the interaction energy of the screw dislocation dissociated in a basal plane
and the two solute atoms can be approximated by
\begin{equation}
	\Delta E(\lambda) = h \left[ -b_i^{(1)}K_{ij}b_j^{(2)} \log{( \lambda / r_{\rm c} )}
	+ \gamma_{\rm Ti} \lambda \right] 
	+ 2 E^{\rm inter}_{\rm SF-X},
	\label{eq:Einter_elas}
\end{equation}
where $\vec{b}^{(1)}$ and $\vec{b}^{(2))}$ are the Burgers vectors of the two partial dislocations
and $K$ the Stroh matrix. For a basal dissociation of the $1/3\,\hkl<1-210>$ screw dislocation, 
an analytical expression of the factor $b_i^{(1)}K_{ij}b_j^{(2)}$ can be found in \cite{Clouet2012}. 
$\gamma_{\rm Ti}=18.7$\,meV/\AA$^2$ is the energy of the basal stacking fault in pure Ti \cite{Kwasniak2016} 
and $h$ is length of the dislocation.
$E^{\rm inter}_{\rm SF-X}$ is the interaction energy of the stacking fault with the solute atom. 
Its calculations is described in \ref{sec:gSF}.
Using for the core radius $r_{\rm c}=a\sqrt{3}/2\sim2.55$\,\AA, the same value as in our previous study \cite{Kwasniak2019}, 
one obtains the results shown as solid lines in Fig. \ref{Fig4}.
With a semi-quantitative agreement with \abinitio calculations, this simple model 
gives a reasonable description of the interaction energy of the solute atoms with the screw dislocation. 
This further confirms that the main effect of the solute atoms is to maintain the basal dissociation 
of the dislocation and that most of the interaction energy 
arises from the energy cost associated with this dissociation 
corrected by the attraction between the solute atoms and the basal stacking fault.
When the interaction energy predicted by Eq. \ref{eq:Einter_elas} becomes positive,
the basal dissociation cannot be anymore stabilized by the solute atoms 
and the interaction energy becomes then null.

\section{Basal glide}

\begin{figure*}[ht!]
\centering
\includegraphics[width=1\textwidth]{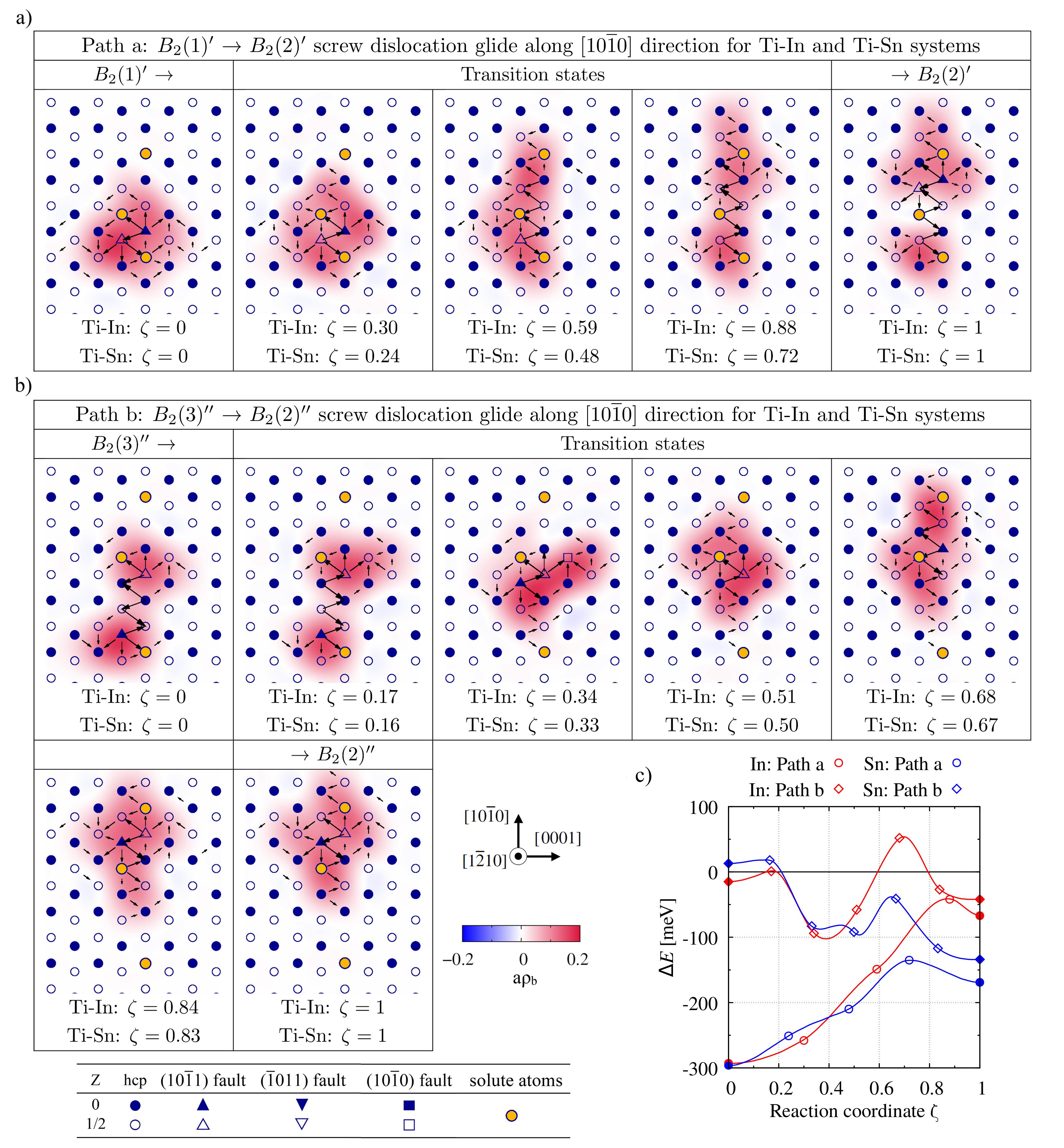}
\caption{(a, b) Two slip paths and (c) the corresponding energy variation for the screw dislocations gliding in a basal plane in Ti-In and Ti-Sn alloys calculated with NEB method. 
Geometry of particular states are shown for Ti-Sn alloy, while equivalent configurations were obtained also for Ti-In system. 
Description of the arrows, symbols and the contour maps is the same as in Fig. \ref{Fig2}. 
The energies of the initial/final structures and intermediate NEB images in (c) are marked by filled and empty circles, respectively.}
\label{Fig5}
\end{figure*}

We now examine if the presence of solute atoms in a basal plane can facilitate basal slip of the screw dislocation.
In that purpose, we determined, with the climbing-image nudged elastic band (NEB) method \cite{Sheppard2008}, the minimum energy path 
between two neighbour positions of the screw dislocation in the \hkl[10-10] direction, 
with the dislocation being partially spread in the basal plane for both the initial and final states.
Two different slip paths, corresponding to two different distributions of solute atoms in the basal slip plane, are considered.
These paths are noted $B_2(1)'$ {\textrightarrow} $B_2(2)'$ and $B_2(3)''$ {\textrightarrow} $B_2(2)''$, 
as their initial and final configurations are similar to the $B_2(1)$, $B_2(2)$ and $B_2(3)$ configurations 
studied in the previous section, but slightly differ because of the presence of a third solute atom in the basal plane (Fig. \ref{Fig5}a and b).
These configurations were selected because of the small separation distance $\lambda$ between the solute atoms, 
leading locally to a high solute concentration.
As shown previously, an attractive interaction between the screw dislocation and the solute atoms is obtained for these configurations (Fig. \ref{Fig4}), 
and the solute atoms promote basal dissociation of the dislocation (Fig. \ref{Fig3}).
The relaxed configurations obtained along these two different basal slip paths are shown for solute Sn in Fig. \ref{Fig5}a and b.
Similar configurations are obtained with the other solute atom In.

Along these paths, an almost purely basal configuration appears as an intermediate transition state. 
This configuration occurs for $\zeta=0.88$ with In (0.72 with Sn) along the path $B_2(1)'$ {\textrightarrow} $B_2(2)'$ (Fig. \ref{Fig5}a)
and for $\zeta=0.68$ with In (0.67 with Sn) along $B_2(3)''$ {\textrightarrow} $B_2(2)''$ (Fig. \ref{Fig5}b). 
In all cases, this configuration corresponds to a saddle point where the energy goes through a maximum (Fig. \ref{Fig5}c),
thus confirming that the basal dissociation of the screw dislocation induced by the solute atoms
is indeed unstable.
For the second considered path (Fig. \ref{Fig5}b),
as the separation distance between the solute atoms is too large to enforce this basal dissociation along all the path, 
the dislocation even goes through an intermediate configuration close to its ground state in pure Ti where the dislocation is dissociated
in a first order pyramidal plane ($\zeta=0.34$ with In and $\zeta=0.33$ with Sn). This intermediate pyramidal configuration corresponds to a minimum
along the path.
Although the dislocation is gliding in a basal plane containing solute atoms stabilizing basal stacking fault, 
the dislocation does not keep a planar configuration dissociated in this basal plane while gliding.

These two different basal slip paths lead to high energy barriers, defined as the difference between the maximum and the minimum energy encountered along these paths:
251 and 146\,meV with In respectively for the paths $B_2(1)'$ {\textrightarrow} $B_2(2)'$ and $B_2(3)''$ {\textrightarrow} $B_2(2)''$,
and 161 and 152\,meV with Sn. 
These energy barriers can be compared to the energy barrier obtained for basal slip in pure Ti,
where, for the same dislocation length, this barrier is only 88\,meV \cite{Kwasniak2019}.
The solute atoms therefore increase the energy barrier for basal slip.  
Basal slip in pure Ti occurs without any basal dislocation, with the dislocation being dissociated either in a pyramidal plane 
or in a prismatic plane while gliding \cite{Kwasniak2019}. 
The fact that solute atoms can induce some basal dissociation of the dislocation does not nevertheless ease basal slip.

Our results therefore point for some hardening of basal slip, both with In and Sn.  
This is consistent with mechanical compression or shear tests performed on single crystals
showing that the critical resolved shear stress for basal slip increases 
with the concentration of simple metal solutes in Ti-Al alloys above a concentration of 1\,at.\% \cite{Sakai1974,Williams2002}. 
This may look surprising as these solute elements are known, on the other hand, 
to increase the activity of basal slip compared to prismatic slip. 
But, the same experiments show that these solute elements also lead 
to a hardening of the prismatic slip system which 
is more pronounced than for basal slip \cite{Sakai1974,Truax1974,Williams2002}, 
thus explaining why such a solute addition decreases the plastic anisotropy.

Post-mortem TEM observations also show 
that basal slip in these Ti alloys is associated with active cross-slip in the prismatic and pyramidal planes
\cite{Sakai1974,Williams2002}.  
This can be rationalized by our $ab$ $initio$ calculations which
reveal that the screw dislocation gliding in a basal plane 
does not retain a configuration dissociated in this basal plane,
even when the concentration in this plane of simple metal atoms is high. 
The gliding dislocation, instead, goes through low-energy configurations 
which are spread in pyramidal or prismatic
and can thus easily cross-slip in these planes. 

\section{Conclusions}

Our \abinitio calculations clearly show that the addition of simple metals like In and Sn
does not manage to fully stabilize a basal dissociation of the $\hkl<a>$ screw dislocation in hcp Ti,
even for the most favorable conditions corresponding to a high concentration of solute atoms in the stacking fault ribbon. 
Although these solute elements greatly reduce the energy cost of the basal stacking fault,
thus corresponding to an attractive interaction with this fault, they do not lead to a planar core exclusively spread in the basal plane
and important spreading in the pyramidal and prismatic planes still exists.
As a consequence of this non planar core, these solute atoms do not ease basal glide. 
Like in pure Ti, basal glide of the \hkl<a> screw dislocation does not process through a basal dissociation of the dislocation,
but involves the low energy configurations which are dissociated in pyramidal and prismatic planes.
The corresponding energy barrier is indeed higher in presence of solute atoms as these atoms act as pinning points.
This is actually in line with the experimental data of Sakai and Fine \cite{Sakai1974} and of Williams \etal \cite{Williams2002} 
who reported an increase,  with Al addition, of the CRSS for basal slip in hcp Ti.
This therefore confirms that the increase of basal slip activity in Ti alloys containing simple metals 
has to be understood as a competition between prismatic and basal slip, with the strengthening effect
of these solute elements being less pronounced for basal than for prismatic slip \cite{Sakai1974,Truax1974,Williams2002}.
Our results show that such a solid solution hardening cannot be simply rationalized through solute interaction 
with the different stacking faults, but that a full modeling of the dislocation is necessary. 

\textbf{Acknowledgments} -
This work supported by the National Science Center under SONATINA 1 project No. 2017/24/C/ST8/00123. 
Computing resources were provided by the HPC facilities of the PL-GRID and CI-TASK infrastructure,
as well as of GENCI-CINES and -TGCC (Grants 2018-096847). 

\appendix
\section{Pseudopotential approximation}
\label{sec:pseudo}

\begin{table}[ht]
\centering
\caption{Comparison of the dislocation energies relative to the pyramidal ground state
and of the Peierls barrier for basal slip in pure Ti 
obtained with $pv$ (10 valence electrons) and $dv$ (4 valence electrons) pseudpotontials. 
Energies are given in meV/{\AA}.}
\label{tab:pseudo}
\begin{tabular}{ p{4.9cm} p{1.8cm} p{0.9cm} }
\hline
Dislocation configuration	& \multicolumn{2}{c}{Pseudopotential} \\
 				& $pv$ & $dv$ \\
\hline
Prismatic 			& 5.6 & 7.5 \\
High energy pyramidal		& 11.9 & 13.1 \\
Saddle point for basal slip	& 13.9 & 15.1 \\
\hline 
\end{tabular}
\end{table}

Poschmann \etal \cite{Poschmann2018} have shown that the energy of the different dislocation configurations 
is sensitive to the number of electrons which are included in the valence shell.
In our calculations, only 4s$^2$ and 3d$^2$ electrons are included in the valence state ($dv$ pseudopotential approximation). 
We have examined how dislocation energy vary in pure Ti when the semi-core 3p$^6$ electrons are also included ($pv$ pseudopotential).
Calculations with both pseudopotentials have been performed keeping the same all other \abinitio parameters
and using exactly the same dislocation setup. 
Results given in table \ref{tab:pseudo} show that the energies of the different metastable configurations of the screw dislocations
vary with the pseudopotential approximation, as well as the Peierls energy barrier for basal glide. 
But this energy variation is small: it is less than 2\,meV/{\AA} and corresponds thus to at most 12\,meV 
for our simulation cell of height $2b$.
Such an energy variation is much smaller than the interaction energy of the dislocation with the solute atoms. 
All \abinitio calculations characterizing dislocation-solute interaction have been therefore performed 
with the $dv$ pseudopotential, as this approximation is much less expensive than with $pv$ pseudopotential.

\section{Variations of solute positions along dislocation line}
\label{sec:fluctuation}

\begin{figure}[th!]
\centering
\includegraphics[width=0.9\columnwidth]{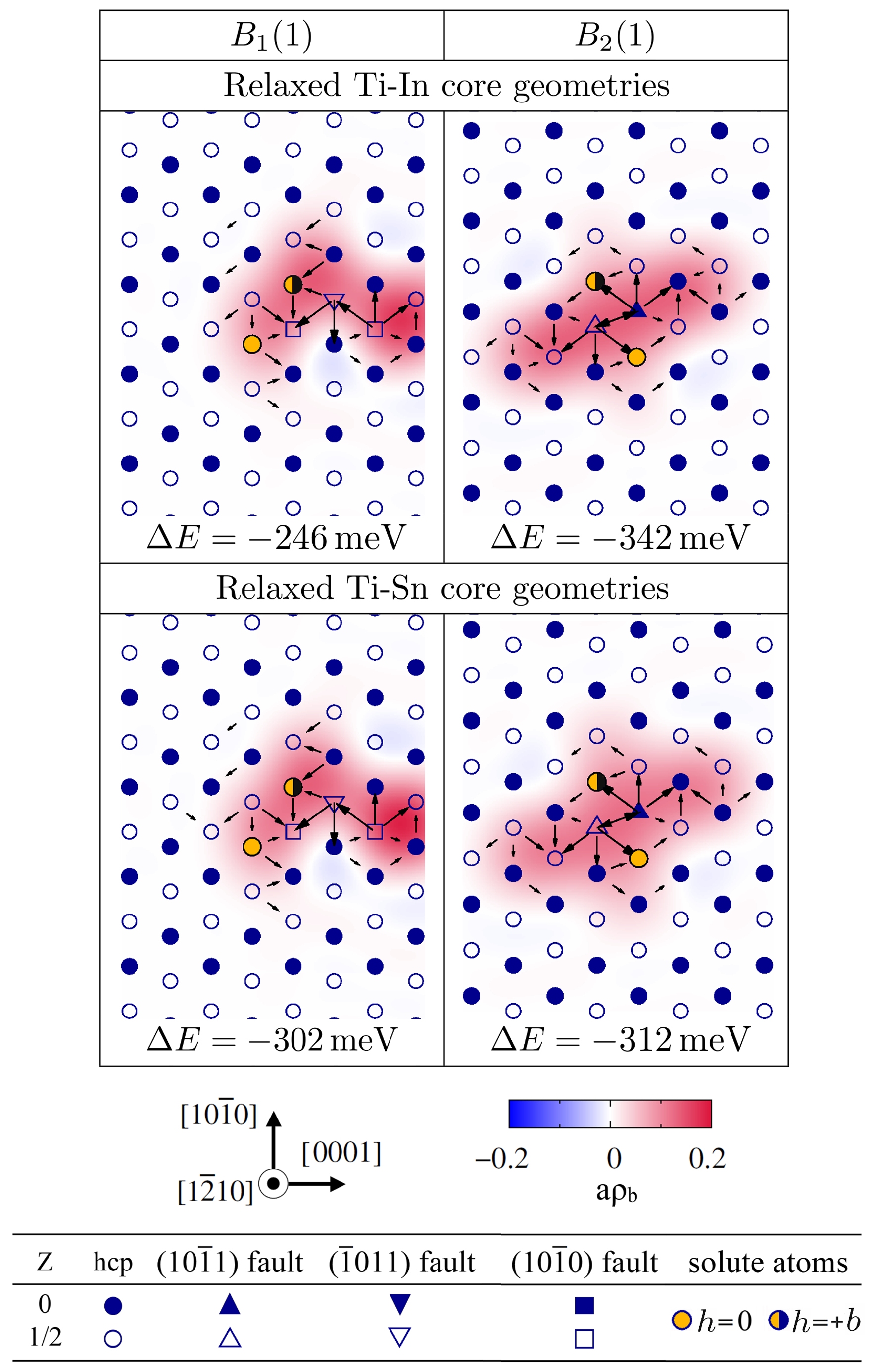}
\caption{Relaxed $B_{1}(1)$ and $B_{2}(1)$ core configurations 
obtained when the solute atoms are introduced with a different relative position along the dislocation line.
$h = 0$ or $+b$ indicates variation of \hkl[1-210] solutes coordinate relative to configurations presented in Figs. \ref{Fig2} and \ref{Fig3}.}
\label{Fig6B}
\end{figure}

The influence of relative solutes position along dislocation line 
was examined for the $B_{1}(1)$ and  $B_{2}(1)$ configurations 
which exhibit the highest absolute value of interaction energy with alloying elements.
The initial configurations of these states were modified 
by displacing one solute atom by a distance $b$ along the \hkl[1-210] direction corresponding to the dislocation line.
The obtained relaxed configurations are shown in Fig.\,\ref{Fig6B} with their corresponding interaction energies.
The same behavior is observed as for the original arrangement where the two solute atoms
are lying in the same \hkl(1-210) plane:
the basal dissociation is unstable and the dislocation core reconstructs and spreads
in pyramidal and prismatic planes, leading to a negative interaction energy. 
For the considered atomic fraction of solutes, the effect of their exact position on the dislocation line
appears therefore to be of secondary importance.

\section{Interaction of basal stacking fault with solute}
\label{sec:gSF}

Following the approach described in Ref. \cite{Kwasniak2016}, 
we have calculated the interaction energy between the basal stacking fault
and a solute atom, In or Sn. 
Calculations have been performed for a stacking of 18 basal planes plus vacuum, 
with each basal plane containing either 4 atoms ($2\times2$ unit cells)
or 6 atoms ($2\times3$ unit cell). 
One Ti atom in the two planes forming the stacking fault is replaced by a solute, 
thus leading locally to a solute concentration of $1/8$ and $1/12$ in the fault.
For Sn, one obtains an interaction energy of $-395$ and $-382$\,meV,
respectively for $1/8$ and $1/12$ solute concentration, 
and for In, $-335$ and $-297$\,meV for the same concentrations.
The average of the two values obtained for each solute has been considered in Eq. \ref{eq:Einter_elas},
\ie $E^{\rm inter}_{\rm SF-Sn}=-389$\,meV and $E^{\rm inter}_{\rm SF-In}=-316$\,meV.

\section*{References}
\bibliography{ref}
\biboptions{sort&compress}

\end{document}